# DENSE RESIDUAL NETWORK FOR RETINAL VESSEL SEGMENTATION

*Changlu Guo[1], Márton Szemenyei[1], Yugen Yi[2*], Ying Xue[3], Wei Zhou[4], Yangyuan Li[1]*

[1] Budapest University of Technology and Economics, Budapest, Hungary
[2]Jiangxi Normal University,Nanchang, China
[3]Eötvös Loránd University, Budapest, Hungary
[4]Shenyang Aerospace University, Shenyang, China

## ABSTRACT

Retinal vessel segmentation plays an imaportant role in the field of retinal image analysis because changes in retinal vascular structure can aid in the diagnosis of diseases such as hypertension and diabetes. In recent research, numerous successful segmentation methods for fundus images have been proposed. But for other retinal imaging modalities, more research is needed to explore vascular extraction. In this work, we propose an efficient method to segment blood vessels in Scanning Laser Ophthalmoscopy (SLO) retinal images. Inspired by U-Net, "feature map reuse" and residual learning, we propose a deep dense residual network structure called DRNet. In DRNet, feature maps of previous blocks are adaptively aggregated into subsequent layers as input, which not only facilitates spatial reconstruction, but also learns more efficiently due to more stable gradients. Furthermore, we introduce DropBlock to alleviate the over-fitting problem of the network. We train and test this model on the recent SLO public dataset. The results show that our method achieves the state-of-the-art performance even without data augmentation.

## 1. INTRODUCTION

Visible changes in retinal vascular structures can indicate many diseases. For example, Diabetic Retinopathy (DR), a complication caused by diabetes, can be diagnosed by changes in the structure of the retinal vasculature [1]. DR is likely to cause blindness [2], which means early detection is vital. Hypertensive Retinopathy (HR) is another retinal disease caused by hypertension [3], and patients with hypertension can find increased vascular curvature or stenosis [4]. Therefore, in medical conditions, certain diseases can be detected and diagnosed by changes in the retina. In obtaining a retinal image the standard image modality is color fundus photography. Another important technique applied to retinal imaging is Scanning Laser Ophthalmoscopy (SLO). The SLO was proposed by [5], which simply scans the fundus with a laser beam. This type of imaging has finer detail, lower exposure, improved contrast compared to color fundus photography, while it also supports direct digital acquisition. However, there are also related problems, such as expensive acquisitions that cannot be performed in true color or sensitivity to motion artifacts.

Segmentation of retinal blood vessels is a particularly important step in current retinal image analysis tasks. However, manual retinal vessel segmentation is an expensive and time-consuming process. Therefore, researchers have proposed numerous automatic segmentation methods to solve this problem [6-8]. However, most prior solutions are designed for standard color retinal fundus images. Since there are some differences between color fundus images and SLO images, these techniques may not properly segment SLO images. Moreover, due to the lack of publicly available datasets, there have been few studies specifically using SLO images [9-12]. A method based on multi-scale matched filter, neural network classifier and hysteresis threshold processing is proposed in [10]. In [9], noise suppression and vascular enhancement are first performed, and then visual features including pixel intensity, wavelet response, and the expansion of Gaussian derivative are calculated, and these features are sent to the neural network to determine the vascular likelihood of each pixel. Zhang et al. [11] proposed an unsupervised method in which an image is back-projected into a 2D space after enhancement of the blood vessel, and then the blood vessel is segmented using global thresholds. In [12], a fully convolutional neural network for bio-image segmentation (U-Net) [13] is used for retinal vessel segmentation, but this method still utilizes the traditional “skip connection" and does not maximize the potential of previous feature map reuse.

Due to their state-of-the-art performance in other fields of computer vision, we used deep neural networks to carry out this research. Inspired by the three well-known networks (U-Net [13], ResNet [14], DenseNet [15]), we propose a

* Corresponding author
This work was supported by the National Natural Science Foundation of China under Grants 61602221, and Chinese Postdoctoral Science Foundation 2019M661117.

Dense Residual Network (DRNet) based on their respective advantages. To take advantage of the limited dataset, we introduce a residual structure to deepen our network to learn the more complex features of retinal images. Although U-Net makes the previous feature mapping reusable through "skip connections", it can effectively solve the problem of spatial information loss. To maximize the potential of feature map reuse, therefore we reuse features via dense connections. In DRNet, the feature maps in the previous blocks are adaptively aggregated into inputs for subsequent layers. Lastly, to reduce the over-fitting problem, we introduce DropBlock [18] to regularize our network more effectively.

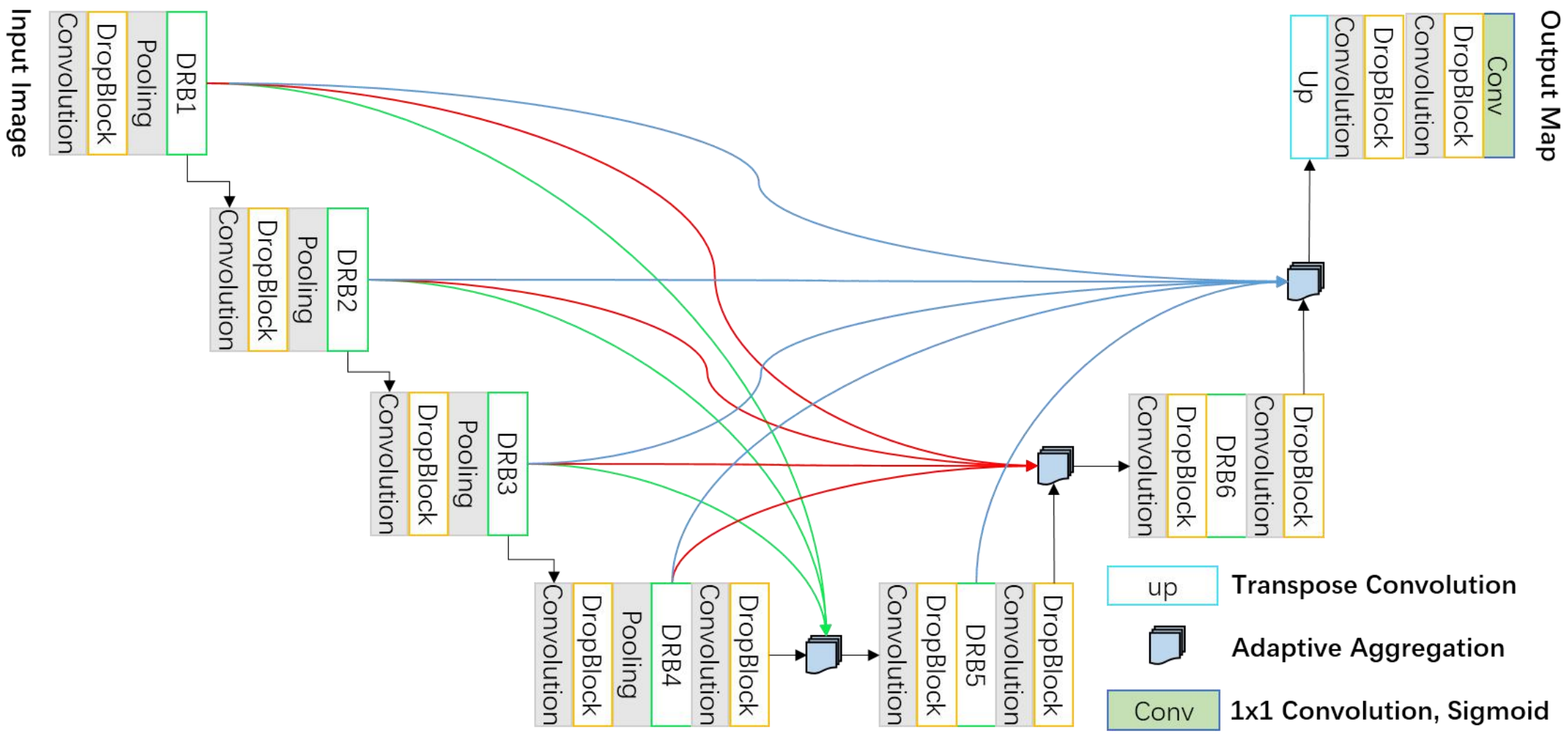

**Fig. 1.** Overview of the proposed Dense Residual Network (DRNet). The output feature maps of the Double Residual Block (from DRB1 to DRB5) are completely reused. The adaptive aggregation module combines the feature maps from all previous blocks to form a new feature map as an input to subsequent blocks. After the aggregation module or DRB, the convolutional layer is used to compress the feature map.

## 2. PROPOSED METHODOLOGY

In this section, we introduce the proposed Dense Residual Network (DRNet) in detail. The full network Architecture is shown in Figure 1. We first detail the Double Residual Block (DRB) in DRNet, and then introduce the Adaptive aggregation of DRB feature maps. Lastly, we briefly introduce DropBlock and elaborate on the details of the network.

### 2.1. Double Residual Block (DRB)

The original structure of our network was modified based on U-Net. Compared with existed methods used the data augmented to reduce overfitting in neural networks, we make full use of the limited annotation data to train a deeper network to extract more complex features in our work. However, simply increasing the number of network layers may hinder training due to the vanishing gradients problem. To solve this problem, we use batch normalization (BN), ReLU activation, 3x3 convolutional layer and DropBlock to construct a residual learning block. As shown in Figure 2, our block is a combination of two residual units, so we named it Double Residual Block (DRB).

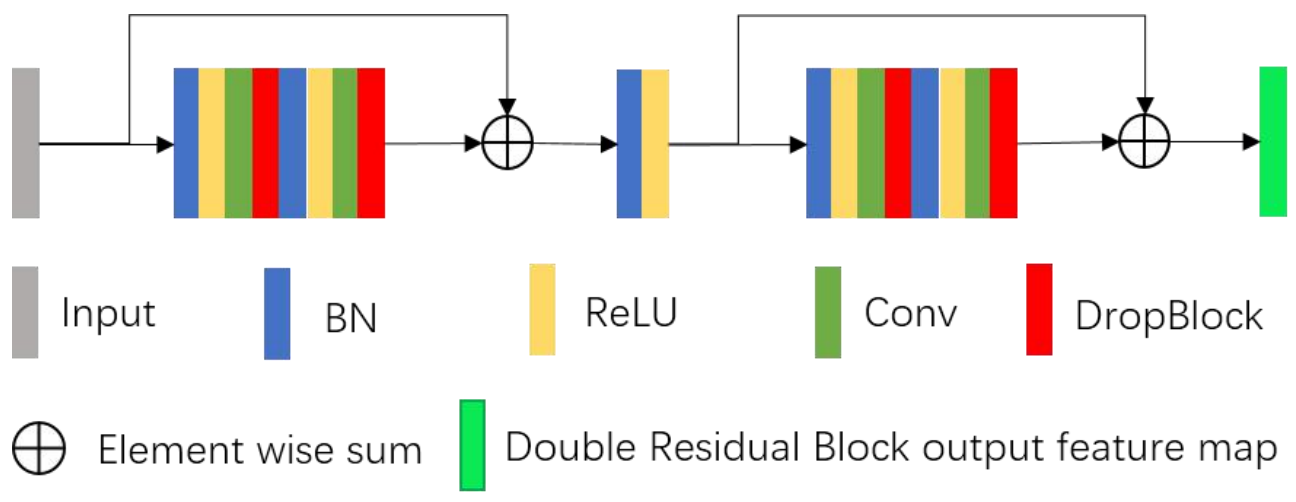

**Fig. 2.** Double Residual Block (DRB), deepen the network through the residual structure.

### 2.2. Adaptive Aggregation of DRB Feature Maps

In previous studies, FCN [16] and U-Net [13] were used to directly reuse previous feature maps through add or concatenate operations. In order to make better use of the feature mapping of the previous blocks, Zhen et al. [17] introduced adaptive aggregation. Similarly, we use the improved adaptive aggregation structure to densely concatenate the feature maps of the previous DRB blocks, as shown in Figure 3.

In DRNet, the high-resolution feature maps of the shallow layers (DRB1, DRB2...) contain coarse semantic information, while the low-resolution features from the deep layers (DRB3, DRB4...) have fine semantic information. The adaptive aggregation structure can fuse all previous DRB feature maps together to generate rich spatial information and context information. The input feature map resolution and feature channel may be different. We first compress the input feature map using a compression layer consisting of DropBlock and $1 \times 1$ convolution operations, except that the directly connected feature map (black arrow) is already compressed, as shown in Figure 3. In order to have the same resolution for all feature maps, we use Max Pooling operation to down-sampling, while up-sampling is done using transposed convolution. Finally, all the resulting feature maps are concatenated into the output for this block.

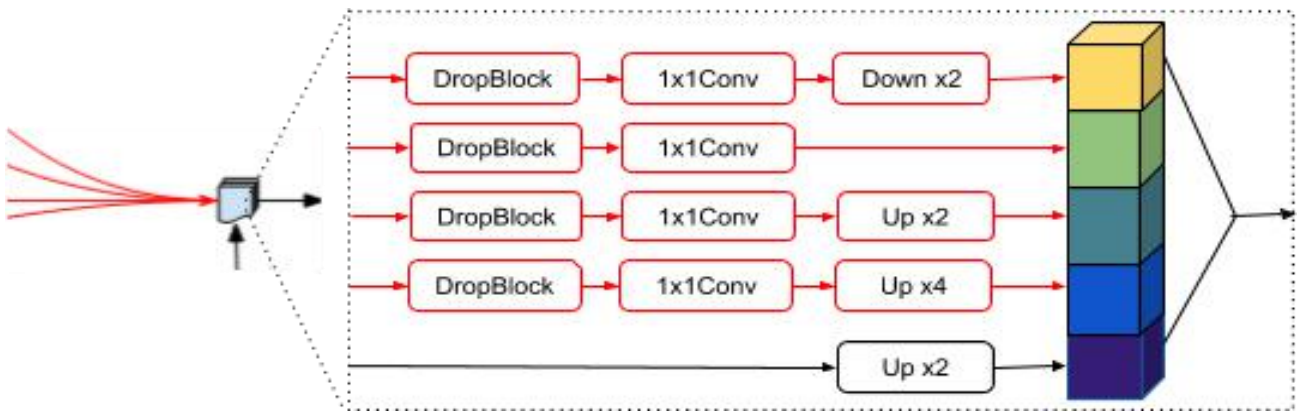

**Fig. 3.** The structure of the adaptive aggregation block. In addition to the directly connected input feature map (black line), all other input feature maps include DropBlock and $1 \times 1$ convolution to adjust the number of feature channels. Then, the size of all the feature maps is made consistent with the output feature map by up-sampling or down-sampling.

### 2.3. DropBlock

According to our initial experiments, overfitting is a serious issue with the original model (without DropBlock [18]), which is likely caused by the increase in depth and the small number of training samples. Dropout is a well-known method for preventing overfitting by randomly dropping activations, although it is mostly used in fully connected layers while it not for convolutional layers because of the spatial correlation among the active cells. However, our model uses a fully convolutional network structure, even with dropout, the semantic information about the input still can be sent to the later layer, causing the network to over-fitting. Intuitively, we need a more structured dropout to regularize our network, so we used DropBlock [18]. The biggest difference compared to dropout is that DropBlock randomly drops continuous areas instead of randomly dropping independent units. This feature allows DropBlock to discard structured semantic information and better prevent overfitting. Ghiasi et al. [18] proposed DropBlock, and also proved its effectiveness in convolutional networks.

### 2.4. Dense Residual Network

The overall architecture of the Dense Residual Network (DRNet) is illustrated in Figure 1. DRNet consists of an encoder part on the left and a decoder part on the right. Each step of the encoder part consists of a $3 \times 3$ convolution operation, a DropBlock, a pooling layer, and a Double Residual Block. Here we set the current number of channels unchanged, until the next step is doubled. In the final step of encoder part, we added a convolution and DropBlock operation and then the output is sent to the first adaptive aggregation operation, which is the beginning of the decoder. The decoder part is more complex, as it consists of three adaptive aggregation operations and two operational blocks consisting of a 3x3 convolutional layer, a DropBlock and a Dense Residual Block. In the final step, we use transposed convolution for up-sampling, followed by two convolution operations with DropBlock, while the final output is computed using $1 \times 1$ convolution and a Sigmoid activation.

## 3. EXPRIMENTS AND RESULTS

In this section, we first introduce the dataset, the hyperparameters used and the evaluation metrics. Then we compare the results of DRNet with the other state-of-the-art methods. We also show a few examples of DRNet's comparison of the segmentation results.

### 3.1. Datasets and Hyperparameters

We used two public retinal datasets, the IOSTAR [19] and the RC-SLO [20] datasets, which were acquired using the Scanning Laser Ophthalmoscopy (SLO) technique. The IOSTAR dataset contains 30 SLO images, each with a resolution of $1024 \times 1024$ pixels. Like [21], we use the first 20 images as the training set and the remaining 10 images as the test set. Due to the small dataset, we randomly selected 10% of the training data set as the validation set. Since the RC-SLO Vascular Patch Dataset contains some difficult conditions, such as background artifacts, bifurcation, crossover, medium cardiovascular reflexes, and high curvature changes, we used it to test the model trained on IOSTAR. The RC-SLO dataset consists of 40 vascular patch images of $360 \times 320$ resolution. To fit our network, we padded each image with zeroes to $1024 \times 1024$ pixels and then cut it to its original size at the time of evaluation.

During training, we set the mini-batch size to 2, the total epochs to 300, and save the parameters with the highest accuracy of the validation set. We set the initial channel number to 16, and Adam as the optimizer with a learning rate of $1 \times 10^{-3}$. The block size of DropBlock in all feature maps is set to 7 and changed to 1 during inference. For best performance, we set the retention probability for each activation unit to 0.86, and like dropout, it is set to 1 during evaluation. All experiments were performed using Keras with Tensorflow as the backend and run on an NVIDIA TITAN XP GPU.

### 3.2. Performance Measurements

To evaluate our model, we used Sensitivity (Sen), Specificity (Spe), Accuracy (Acc), Area Under the ROC Curve (AUC) and Matthews Correlation Coefficient (MCC) as evaluation metrics. The aforementioned metrics are calculated as follows:

$$\text{Sen} = \frac{TP}{TP + FN}, \quad \text{Spe} = \frac{TN}{TN + FP}, \quad \text{Acc} = \frac{TP + TN}{N}$$

$$\text{MCC} = \frac{TP \times TN - FP \times FN}{\sqrt{(TP + FP) \times (TP + FN) \times (TN + FP) \times (TN + FN)}}$$

where TP is true positive area, which means that the blood vessel area in the ground truth is correctly classified as blood vessel in the segmentation result. Conversely, if the region is classified incorrectly in the segmentation result, it is called false negative (FN). True negatives (TN) indicate that the non-vascular regions of ground truth are correctly classified as non-vascular in the segmentation results. If the background region is classified incorrectly in the segmentation result, it is called false positive (FP). Also, N=TP+TN+FP+FN. The area under the ROC curve (AUC) can be used to measure the performance of the segmentation. If the AUC value is 1, it means perfect segmentation.

**Table 1**: COMPARISONS WITH EXISTING METHODS ON **IOSTAR** DATASET

| Method | Year | Sen | Spe | Acc | AUC | MCC |
|---|---|---|---|---|---|---|
| Abbasi-Sureshjani et al. [9] | 2015 | 0.7863 | 0.9747 | 0.9501 | 0.9615 | 0.7752 |
| Zhang et al. [11] | 2016 | 0.7545 | 0.9740 | 0.9514 | 0.9626 | 0.7318 |
| Meyer et al. [12] | 2017 | 0.8038 | 0.9801 | 0.9695 | 0.9771 | 0.7920 |
| Srinidhi et al. [22] | 2018 | **0.8269** | 0.9669 | 0.9564 | 0.9663 | 0.7057 |
| **DRNet** | **2019** | 0.8082 | **0.9854** | **0.9713** | **0.9873** | **0.8017** |

**Table 2**: COMPARISONS WITH EXISTING METHODS ON **RC-SLO** DATASET

| Method | Year | Sen | Spe | Acc | AUC | MCC |
|---|---|---|---|---|---|---|
| Zhang et al. [11] | 2016 | 0.7787 | 0.9710 | 0.9512 | 0.9626 | 0.7327 |
| Meyer et al. [12] | 2017 | 0.8090 | 0.9801 | 0.9623 | 0.9807 | 0.7905 |
| Srinidhi et al. [22] | 2018 | **0.8488** | 0.9666 | 0.9581 | 0.9678 | 0.7029 |
| **DRNet** | **2019** | 0.8151 | **0.9879** | **0.9744** | **0.9848** | **0.8190** |

### 3.3. Segmentation Results and Evaluation

We first evaluate our model at IOSTAR, then on the RC-SLO dataset. Tables 1 and 2 summarize the results of each metric by comparing our method with the results of other state-of-the-art methods on the same dataset. Most of the metrics indicate that the proposed method surpasses the previous state-of-the-art results. Specifically, DRNet achieves the highest MCC (i.e. 0.97%/2.85% higher than the second best), the highest AUC (1.02%/0.41% higher than the second best), the highest accuracy (0.18%/1.21% higher than the second best), the highest specificity and comparable specificity on both datasets. The above results clearly show that DRNet is an effective method for retinal vessel segmentation of Scanning Laser Ophthalmoscopy images. Figures 4 and 5 show sample results of the proposed method for IOSTAR and RC-SLO images, respectively.

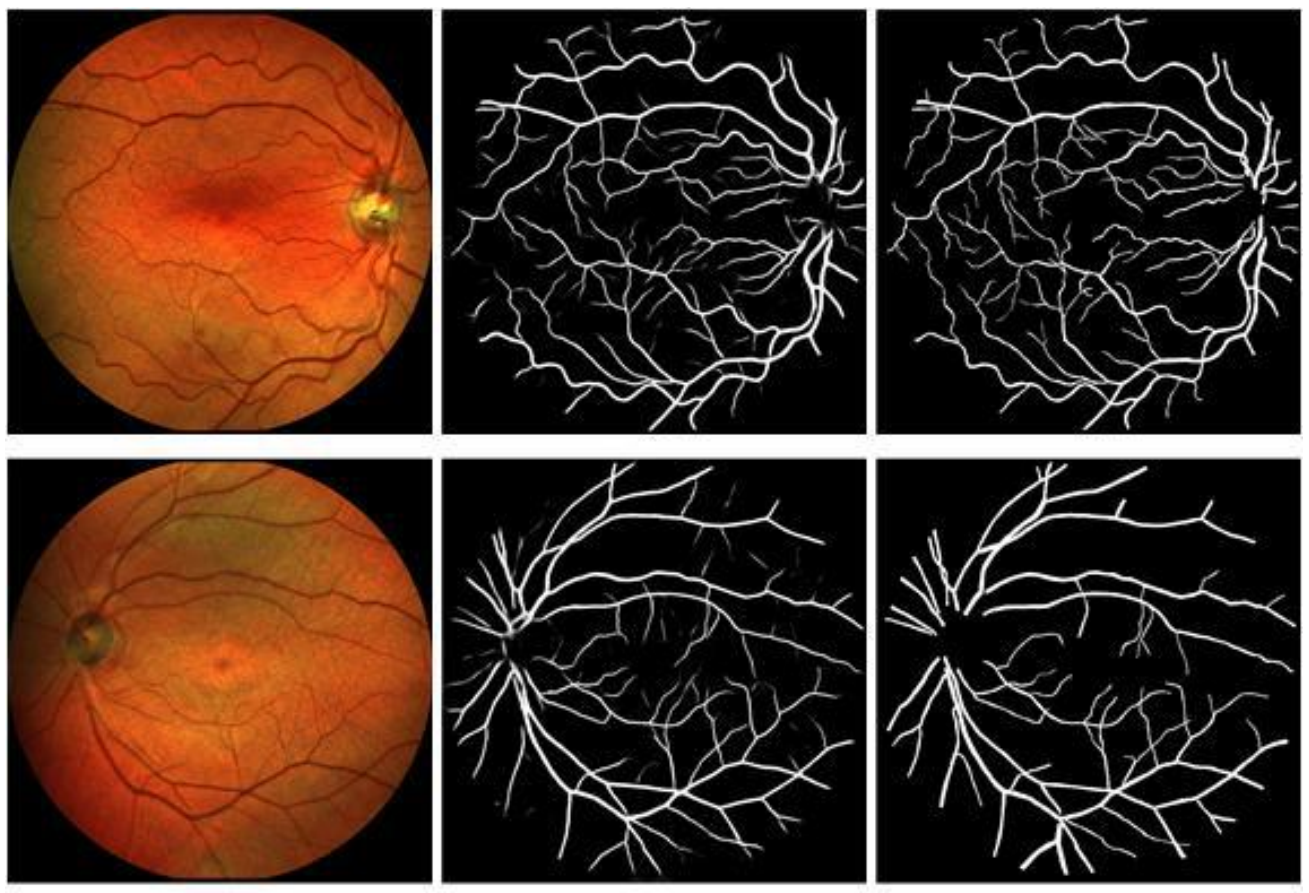

(a) (b) (c)

**Fig. 4.** Segmentation results for IOSTAR dataset. (a) Real image. (b) Segmentation results. (c) Ground truth annotation.

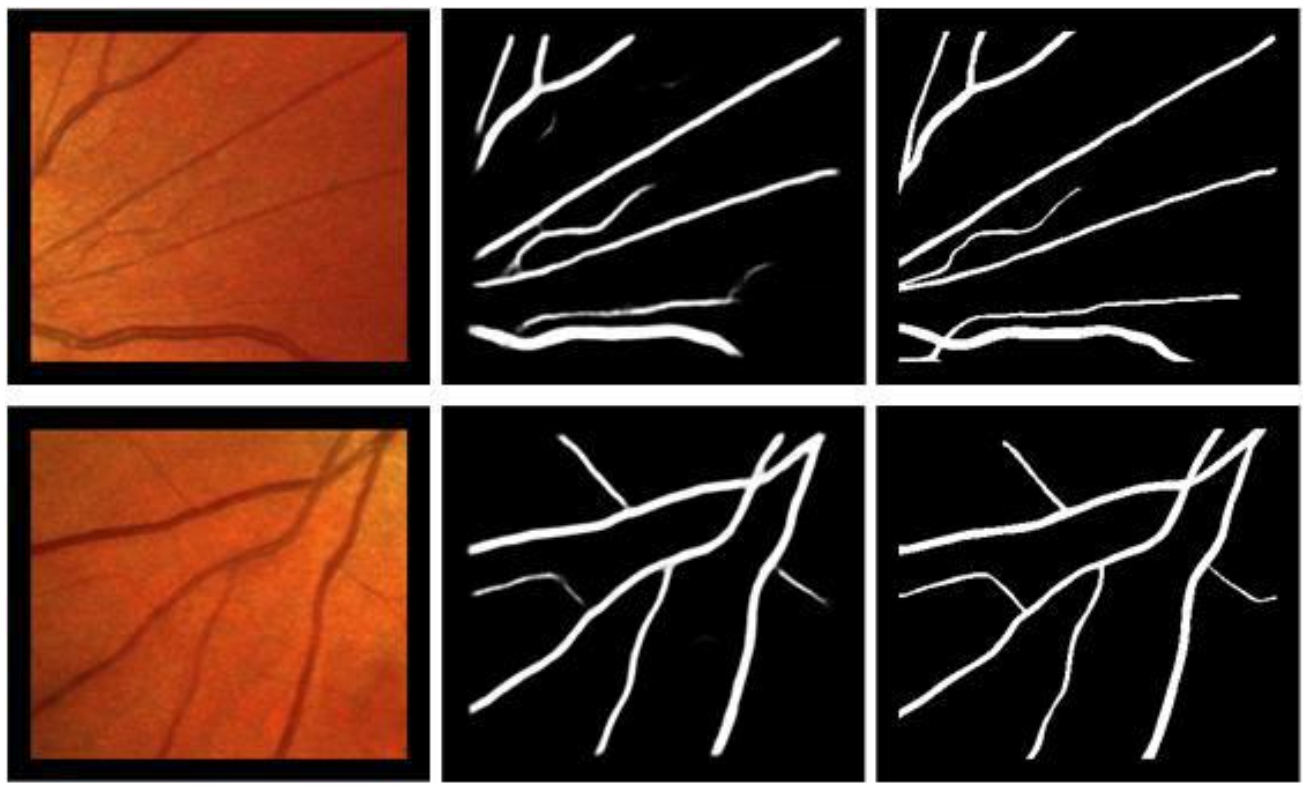

(a) (b) (c)

**Fig. 5.** Segmentation results for RC-SLO dataset. (a) Real image. (b) Segmentation results. (c) Ground truth annotation.

## 4. CONCLUSION

In this paper, we proposed a deep dense residual network (DRNet) for retinal vessel segmentation of Scanning Laser Ophthalmoscopy images, which uses adaptive aggregation structures to connect feature maps from previous Double Residual Blocks (DRB) to subsequent steps. The dense connections make DRNet effective in retinal vessel segmentation because DRNet like ResNet and Dense Net can reduce the vanishing gradient problem and learn more effectively. In addition, the network also uses DropBlock to alleviate overfitting. Experiments have shown that even without any data augmentation, our model is superior to previous networks in two retinal datasets using Scanning Laser Ophthalmoscopy.